\documentclass[aps,pre,twocolumn,showpacs]{revtex4}
\usepackage{epsfig}
\usepackage{times}
\begin{document}

\title{Segregation process and phase transition in cyclic predator-prey models \\
with even number of species}

\author{Gy\"orgy Szab\'o$^1$, Attila Szolnoki$^1$, and Gustavo Ariel Sznaider$^2$}
\affiliation
{$^1$Research Institute for Technical Physics and Materials Science
P.O. Box 49, H-1525 Budapest, Hungary  \\
$^2$Quantitative Applied Methods,
Faculty of Agronomy, University of Buenos Aires,
Av. San Martin 4453, Buenos Aires (1417), Argentina}

\begin{abstract}
We study a spatial cyclic predator-prey model with an even number of species (for $n=4$, 6, and 8) that allows the formation of two defective alliances consisting of the even and odd label species. The species are distributed on the sites of a square lattice. The evolution of spatial distribution is governed by iteration of two elementary processes on neighboring sites chosen randomly: if the sites are occupied by a predator-prey pair then the predator invades the prey's site; otherwise the species exchange their site with a probability $X$. For low $X$ values a self-organizing pattern is maintained by cyclic invasions. If $X$ exceeds a threshold value then two types of domains grow up that formed by the odd and even label species, respectively. Monte Carlo simulations indicate the blocking of this segregation process within a range of $X$ for $n=8$.      
\end{abstract}


\maketitle

Interesting phenomena in the cyclic predator-prey systems initiated a progressive research in the last decades. In the simplest case the system has three species dominating cyclically each other, that is, $S_1$ beats $S_2$ beats $S_3$ beats $S_1$. Cyclic dominance occurs in many systems including biological models \cite{kerr_n02,durrett_tpb98,johnson_prslb02,mobilia_jsp07} and evolutionary games \cite{nowak_jtb90,traulsen_pre04}. It is well known that such a cyclic dominance can sustain all of the three species in a well-mixed community \cite{may_siam75,hofbauer_98}. In the spatial version of this system the individuals stay on the sites of a lattice \cite{tainaka_prl89} and the cyclic invasions maintain a self-organizing pattern in the spatial distribution of species. Nowadays, this system is frequently referred to as an evolutionary spatial Rock-Scissors-Paper game (for a review see \cite{szabo_pr07}). 

Within the framework of mean-field approximation (assuming well-mixed population) the species densities can be either stationary or oscillatory (for the latter case both the sum and product of species densities are conserved quantities) \cite{itoh_ptp87,hofbauer_98}. Furthermore, these systems exhibit unusual response to the variation of invasion rates or to any external support \cite{tainaka_pla93,frean_prsb01}. For finite population the system evolves into one of the homogeneous state via a Moran process \cite{itoh_pla94,mobilia_pre06}. On the other hand, when the species are distributed on a lattice the cyclic invasions yield a self-organizing pattern \cite{tainaka_pre94,kerr_n02} providing stability against some types of external invaders \cite{boerlijst_pd91,szabo_pre01a} if the spatial dimension is larger than one ($d>1$). For the one-dimensional lattice the dynamical rule results in growing domains and super-domains \cite{tainaka_prl89,frachebourg_prl96,frachebourg_pre96}. On the Bethe lattice one can observe limit cycle behavior (for a degree of $z=3$ or $4$) and the oscillation grows until the system reaches one of the homogeneous states if $z \ge 6$ \cite{sato_mmit97,szolnoki_pre04a}. More complicated behavior is reported on some types of directed graphs \cite{ying_jpa07}.

Starting from the simplest model there are many ways for the generalization of the evolutionary Rock-Scissors-Paper game. A straightforward possibility is to increase the number $n$ of species (states) so that the cyclic invasion remains valid, i.e., $S_1$ invades $S_2$ invades $S_3$ {\it etc.} and finally $S_n$ invades $S_1$. In the one-dimensional lattice domain growth can be observed until $n \le 5$, otherwise the spatial distribution tends towards a frozen domain structure where the species staying in neighboring domains cannot invade each other (shortly, they are neutral). Similar fixation process was reported for both the two- and three-dimensional lattices if $n$ exceeds a critical value dependent on $d$ \cite{frachebourg_jpa98}. Further relevant observation is related to the parity of $n$ affecting some features (e.g., the sensitivity to the inhomogeneous invasion rates) in the system \cite{itoh_ptp87,sato_amc02}. Consequences of supplementary microscopic processes are also investigated. For example, one can introduce mutation, extinction, and empty sites to which the neighboring species can jump, allowing spatial mixing \cite{szabo_pre04a,he_ijmpc05}. Now our efforts will be concentrated on a lattice Lotka-Volterra system with an even number $n$ of species dominating cyclically each other while the stochastic local mixing is described by a site exchange mechanism between neutral pairs. The simplest model where such a process can be introduced is the four-species version that exhibits a phase transition when increasing the strength of mixing \cite{szabo_pre04a,szabo_jpa05}. More precisely, for low rate of mixing one can observe a self-organizing pattern resembling the sample of evolutionary Rock-Scissors-Paper game. On the contrary, if the strength of mixing exceeds a threshold value then phase segregation occurs, i.e., the well-mixed odd end even label species form growing domains and finally one of these states dominates the whole finite system. Our primary interest was to investigate how the critical value of mixing decreases when increasing the number of species. During this study, however, an unexpected intermediate phase was found for $n =8$ and $10$ as will be reported below.

We consider a lattice Lotka-Volterra model on a square lattice where each site $x$ is occupied by a single individual belonging to one of the $n$ species ($n$ is even), that is, the distribution of species can be described by the site variables $s_x=1, \ldots, n$ referring to the label of species. The time evolution of the species distribution is determined by invasions between the site $x$ (chosen at random) and one of the (randomly chosen) neighboring sites $y$ if the sites are occupied by a predator-prey pair, i.e., the $(s_x,s_y)$ [and also the $(s_y,s_x$)] pair transforms into $(s_x,s_x)$ if the species $s_x$ is the predator of $s_y$. The predator-prey relation is defined by a cyclic food web and the invasion rates between any predator-prey pair are equivalent and chosen to be unity. Besides it, if the species $s_x$ and $s_y$ are neutral then the species may exchange their site with a probability $X$ characterizing the strength of mixing. Obviously, nothing happens if $s_x=s_y$. During the time unit (called Monte Carlo step, shortly MCS) the above elementary process is repeated once on average for each site. 

For the Monte Carlo (MC) simulations the system is started from a random initial state (providing equivalent (average) number of individuals for the species) on a square lattice (consisting of $N=L \times L$ sites) with periodic boundary conditions. When repeating the above defined elementary step (invasion or local mixing) for sufficiently large size the system evolves into a stationary state that can be described by the average density $\rho_i$ of species $i$ (satisfying the condition $\sum_{i=1}^{n} \rho_i =1$) and by the pair configuration probabilities $p_2(i,j)$ of finding species $i$ and $j$ on two neighboring sites. In the limit $N \to \infty$ each species density remains constant, that is $\rho_i(t)=1/n$. Whereas the ordering processes can be well quantified by considering the time-dependence of pair configuration probabilities. Due to the symmetries we will distinguish two basic quantities. The predator-prey pair probability is given as
\begin{equation}
p_{pp}(t) = \sum_{i=1}^{n} [p_2(i,i+1)+p_2(i+1,i)]
\label{eq:ppp}
\end{equation}
where the time-dependence of $p_2(i,j)$ is not denoted and $i+1$ means $1$ for $i=n$. The other analogous quantity is the neutral pair probability that can be expressed as
\begin{equation}
p_{n}(t) = \sum_{i=1}^{n} \sum_{k=2}^{n-2} p_2(i,i+k) \;
\label{eq:np}
\end{equation}
where $i+k$ is cyclically reduced to the range 1 to $n$. For the MC simulations these quantities are determined by averaging over a suitable sampling interval.

For small size (e.g., $L < 10$) the system quickly develops into one of the states in which several species and simultaneously the predator-prey invasions are missing, thus the composition remains constant. The average fixation (transient) time increases fast with the linear size $L$. For sufficiently large size, however, most of the latter phases can also be observed locally within small patches and the evolution of spatial distribution is affected by the competition between these phases \cite{szabo_jpa05,szabo_pr07}.

\begin{figure}[ht]
\centerline{\epsfig{file=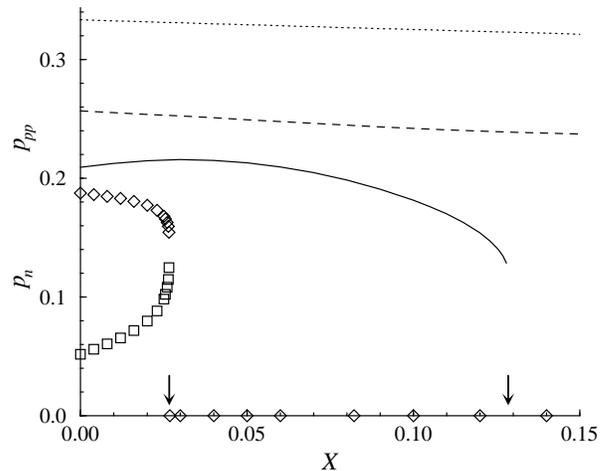,width=8cm}}
\caption{\label{fig:lv4scx}MC data for the predator-prey pair $p_{pp}$ (diamonds) and neutral pair $p_{n}$ (squares) probabilities as a function of $X$ in the final stationary state for $n=4$. Dotted, dashed, and solid lines represent the prediction of generalized mean-field approximation for $p_{pp}$ at the levels of $2 \times 1$-, $2 \times 2$-, and $3 \times 3$-site clusters. The left arrow indicates the position of $X_c$, the right arrow shows the corresponding prediction obtained by the $3 \times 3$-site approximation. }
\end{figure}

First we briefly recall the MC results obtained for $n=4$ \cite{szabo_jpa05}. According to the simulations all the four species are sustained by the cyclic invasions if $X<X_c(4)=0.02662(2)$. Within this region of $X$ the value of $p_{n}$ increases monotonously with $X$ meanwhile an opposite trend occurs in $p_{pp}$ as demonstrated in Fig.~\ref{fig:lv4scx}. At the critical value of $X$ a sudden change occurs in both quantities because, through a domain growing process, the finite system evolves into a state where either the odd or the even labelled species form a well-mixed phase, that is, $\rho_1(\infty)=\rho_3(\infty)=1/2$ and $\rho_2(\infty)=\rho_4(\infty)=0$ or $\rho_1(\infty)=\rho_3(\infty)=0$ and $\rho_2(\infty)=\rho_4(\infty)=1/2$. In both phases $p_{pp}(\infty)=0$ and $p_{n}(\infty)=1/2$ (if $X>X_c(4)$). 

To support our previous MC results we have also performed generalized mean-field approximations at different levels. In the case of the four-species version of this model, all the configuration probabilities on $2 \times 1$-, $2\times 2$-, and $3 \times 3$-site clusters have been evaluated (for details of the method see \cite{szabo_pr07}). Evidently, the traditional mean-field approach (one-site approximation) cannot take into account the effect of mixing. At the levels of $2 \times 1$- and $2 \times 2$-site cluster approximation this method is not capable to describe the transition observed by MC simulations although the solution of many well-mixed phases exists. The more accurate $3 \times 3$-site approximation predicts a phase transition at $X=X_c^{(9s)}(4)=0.1285(5)$. The large deviation from the MC results indicates  the importance of consecutive elementary steps yielding important correlations in the spatial distribution.  

For sufficiently strong mixing the phase segregation process seems to be a robust phenomenon as it is observed for other dynamical rules \cite{szabo_pre04a,he_ijmpc05,szabo_pr07}. It turned out that the well-mixed phases of the odd (as well as even) label species can be considered as a defensive alliance because within this spatial association the species guard each other against the external invaders. For example, if species 1 is attacked by an individual of species 4 then one of the neighboring species 3 strikes back within a short time. Due to the cyclic symmetry species 1 guards species 3 against species 2 and similar mechanism protects the well mixed spatial association of the even label species.  
 
\begin{figure}[ht]
\centerline{\epsfig{file=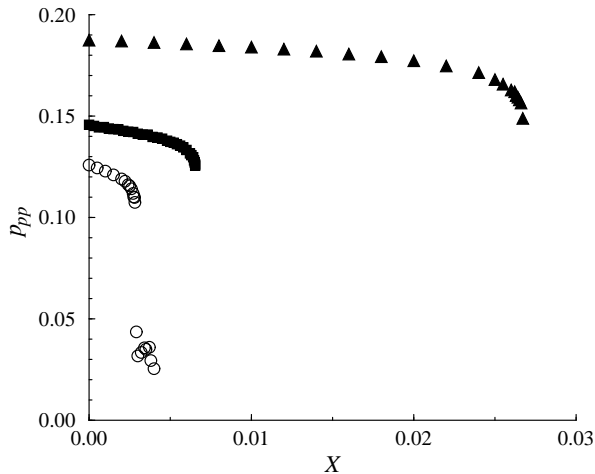,width=8cm}}
\caption{\label{fig:lv468xpp}The non-vanishing predator-prey pair probabilities {\it vs.} $X$ in the stationary states for $n=4$ (triangles), $6$ (squares), and  $8$ (circles).}
\end{figure}

One can easily check that the above concept of defensive alliances remains valid for any even number of species. Namely, within the well-mixed phase of the odd label species the species protect each other against the invasion of even label species and {\it vice versa}. Thus for larger $n$ one can expect a similar phase transition from the cyclic self-organizing pattern to the phase segregation phenomenon if we increase the value of $X$. MC simulations confirm this expectation for $n=6$ as illustrated in Fig.~\ref{fig:lv468xpp}. The predator-prey probability ($p_{pp}$) drops suddenly to zero at $X=X_c(6)=0.00654(2)$. The generalized mean-field analysis of this system is not performed for $n \ge 6$ because the numerical solution becomes time-consuming due to the large number of configurations at a sufficiently accurate level (e.g., $n^9$ configurations exist on a $3 \times 3$ cluster). 

Notice that many possible solutions emerge if $n$ is increased. For example, the well-mixed phases of the mutually neutral species (e.g., 1+3+6 for $n=8$) with arbitrary composition are stationary states. All these states can occur in the MC simulations for small sizes ($L \le 10$) and can also be reproduced by the generalized mean-field methods. Indeed, the solutions of the subsystems (several species are missing) are solutions for the whole system too. Despite of the large number of possible solutions, the visualization of spatial distributions in the MC simulation indicates the presence of only the above mentioned relevant phases for sufficiently long times.  

Surprisingly, for the eight-species system three types of phases (behaviors) can be distinguished as illustrated in Fig.~\ref{fig:lv468xpp}. These MC data are obtained for $L=400$ or $600$. For these sizes the domain growing process ends within a few million MCS and one of the four-species defensive alliances (consisting of only the odd or even label species) prevails the whole system in the final state if $X>X_{c2}(8)\simeq 0.0042(5)$. The self-organizing spatio-temporal pattern can be maintained by the cyclic invasions until $X<X_{c1}(8) \simeq 0.00285(3)$. As a consequence, the MC simulations reveal the appearance of an intermediate phase within a region of $X_{c1}(8)<X<X_{c2}(8)$ where the domain growing process stops (or becomes extremely slow).

\begin{figure}[ht]
\centerline{\epsfig{file=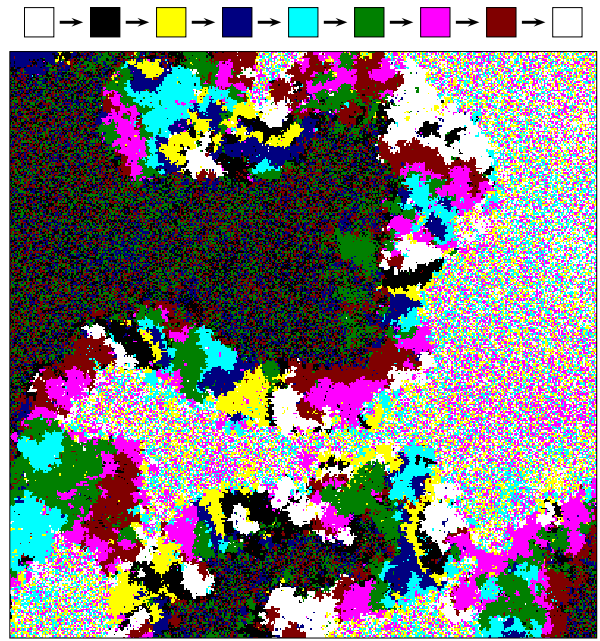,width=8cm}}
\caption{\label{fig:s8grdoms}(Color online) Spatial distribution of species after $400\,000$ MCS for $X=0.003$ if initially the eight species were distributed randomly on a square lattice. The cyclic dominance between the eight (colored) species is indicated at the top and the snapshot shows a $400 \times 400$ portion of the whole system with a size of $1600 \times 1600$.}
\end{figure}

To illustrate the formation of defensive alliance in the intermediate region ($X_{c1}<X<X_{c2}$) the odd and even label species are denoted by light and dark colors in Fig.~\ref{fig:s8grdoms}. The snapshot for $t=400\,000$ MCS shows clearly that the territories of defensive alliances are separated by a boundary layer where the cyclic invasions govern the time evolution. Similar patterns (with a thickness dependent on $X$) can be observed during the domain growing process for $X > X_{c2}(8)$. We have to emphasize that these boundary layers play crucial role in the formation of final pattern. Namely, these layers serve as a symmetric species reservoir for both defensive alliances and help the equalization of their composition via diffusion. 

As mentioned above the domain growing process can be investigated quantitatively by recording the probability of predator-prey pairs because such a constellation occurs exclusively within the boundary layers. One can think that the inverse of $p_{pp}$ is proportional to the average linear size of domains of defensive alliances.

\begin{figure}[ht]
\centerline{\epsfig{file=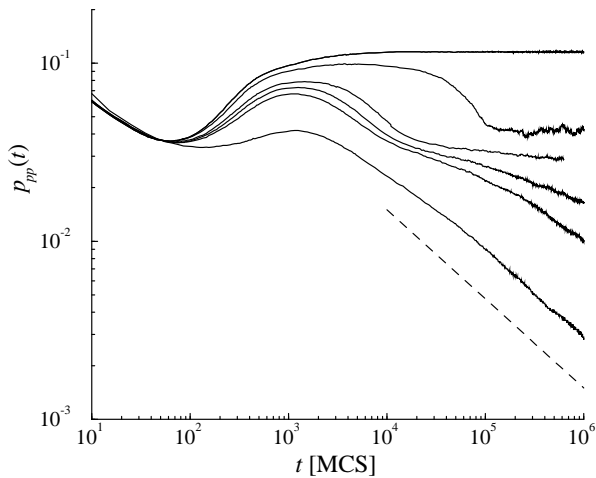,width=8cm}}
\caption{\label{fig:s8pinvt}Log-log plot for the time-dependence of the predator-prey pair probabilities in the eight-species system for $X=0.0025$, 0.003, 0.004, 0.0045, 0.005, and 0.008 (from top to bottom). The MC results are obtained on a square lattice with a linear size of $L=2800$ and the plotted data are smoothed by averaging over a time-window (typically $\Delta t\simeq t/40$ maximum 1000 MCS). Dashed line shows the slope of $-1/2$ characterizing domain growth driven by the decrease of interfacial energy \cite{bray_ap94}.}
\end{figure}

Figure \ref{fig:s8pinvt} shows some typical behaviors when considering the time-dependence of $p_{pp}$ for such a system size where $L$ is significantly larger than the average domain size at the end of simulation (here at $t= 10^6$ MCS). In order to suppress the short-time fluctuations, the numerical data of $p_{pp}(t)$ are averaged over a time interval with a typical width of $t_w \simeq \min(t/20, 1000 \mbox{ MCS})$. The upper curve ($X=0.0025$) in Fig.~\ref{fig:s8pinvt} illustrates that the frequency of invasions reaches a high stationary value ($p_{pp} \simeq 0.11$) characterizing the cyclic self-organizing pattern if $X < X_{c1}(8)$. The lowest curve (for $X=0.008$) indicates a typical domain growing process when the asymptotic behavior ($p_{pp} \sim t^{-1/2}$) becomes similar to phase ordering process with non-conserved dynamics \cite{bray_ap94}. From the plotted MC data at $X=0.005$ and $0.008$ one can suggest similar asymptotic behavior. Data for $X=0.003$, however, indicates that the domain growth is stopped and the process itself is resembling the segregation of a water-oil mixture in the presence of surfactant \cite{gompper_94,henriksen_pre00}. Notice that in this spatio-temporal pattern all the eight species remain alive, thus the spontaneous formation of this inhomogeneous pattern exemplifies a way how the biodiversity can be maintained for a long time. We have to emphasize that, in the absence of a clear theoretical explanation of this phenomenon, we cannot exclude the appearance of a slower domain growing process for a longer time scale ($t > 10^6$ MCS).  

\begin{figure}[ht]
\centerline{\epsfig{file=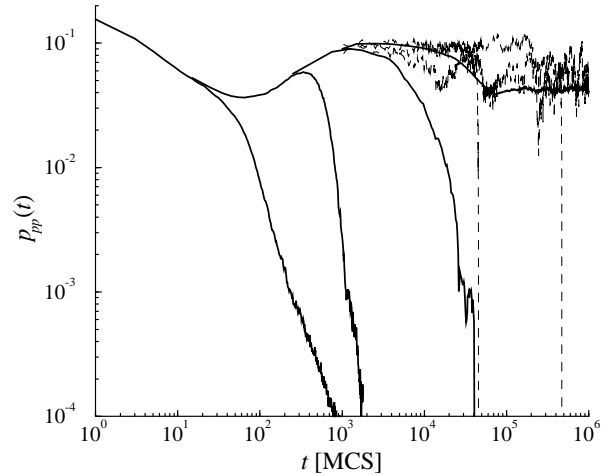,width=8cm}}
\caption{\label{fig:sizeff}MC data of $p_{pp}(t)$ for $X=0.003$ when varying the system size. Thick lines from left to right show the results for $L=10$, $30$, and $100$ after averaging over $10^4$, $10^3$, and $10^2$ runs, respectively. To demonstrate the large fluctuations in the fixation time the subsequent three dashed lines illustrate the results of three runs for $L=200$. The last thick line (representing a single run for $L=1000$) resembles data (obtained for $L=2800$) plotted in Fig.~\ref{fig:s8pinvt}.}
\end{figure}

Within the intermediate region of $X$ the reproduction of numerical data is poor (see Fig.~\ref{fig:lv468xpp}) because the system behavior is perturbed by large and slow fluctuations (see data for $X=0.003$ in Fig.~\ref{fig:s8pinvt}). To have a deeper insight into this ordering process the function $p_{pp}(t)$ is plotted in Fig.~\ref{fig:sizeff} for different system sizes.   

Figures \ref{fig:s8pinvt} and \ref{fig:sizeff} indicate clearly that the sufficiently large domains of the well mixed phases of odd and even label species are formed after a transient time as long as 20,000 MCS. The symmetric composition of these domains (namely, $\rho_0=\rho_2=\rho_4=\rho_6=1/4$) is ensured by the long contact (interaction) with the boundary layers serving as symmetric species reservoirs \cite{szabo_pre04a}. At the same time, the final spatial structure of the boundary layers is also affected by their interactions with the well-mixed phases of neutral species. These mutual effects can be a cause of the blocking of domain growing process in the intermediate region of $X$. The mentioned process is observed if the system size $L$ exceeds significantly the typical domain size ($l \simeq 100$ lattice unit in Fig.~\ref{fig:s8grdoms}). In the opposite case this proper structure cannot build up because the system develops into a state consisting of two, three or four neutral species. For example, the left curve ($L=10$) in Fig.~\ref{fig:sizeff} represents a fast evolution into a final state where about one third (two percent) of runs end with three (two) neutral species. Although the probability of finding four neutral surviving species increases with system size, the composition of the final state is far from being symmetric if $l \alt L$.  

Due to the extremely long transient times at the boundaries of the intermediate region, more accurate determination of the critical values and systematic analysis of the corresponding phase transitions exceed our computing capacity. 

The preliminary results indicate similar behavior for $n=10$. In this case the system exhibits longer relaxation (related to the slower formation of the corresponding defensive alliances) making the rigorous analysis more difficult. 

In summary, the present work is focused on the spatial formation of two defensive alliances on a two-dimensional lattice Lotka-Volterra model with even number $n$ of species invading cyclically each other with the same rate. The introduction of local mixing (with strength characterized by the site exchange probability $X$ between the neutral species residing on neighboring sites) supports the formation of the well-mixed distribution of odd or even label species representing two equivalent defensive alliances. Phase segregation process is observed if the mixing rate exceeds a threshold value dependent on $n$. According to our MC simulations this system evolves into a pattern (for $n \ge 6$) where the domains of defensive alliances are separated by boundary layers having a different structure. This phenomenon raises further questions about the role of boundary layers in these types of complex systems.

\begin{acknowledgments}

This work was supported by the Hungarian National Research Fund
(Grant No. T-47003).

\end{acknowledgments}


\end{document}